\documentclass[iop,twocolumn,numberedappendix]{aastex62}

\newcommand{\arw}{$\rightarrow$}
\newcommand{\water}{H$_2$O}
\newcommand{\htop}{H$_3$O$^+$}
\newcommand{\waterg}{H$_2$O$_{\rm(gr)}$}

\newcommand{\HH}{H$_2$}

\newcommand{\waterp}{H$_2$O$^+$}
\newcommand{\Hp}{H$^+$}

\newcommand{\hcop}{HCO$^+$}
\newcommand{\uvphoton}{$\gamma_{\rm UV}$}
\newcommand{\xrayphoton}{$\gamma_{\rm UV, fl}$}

\newcommand{\density}{\,$\rm g\,cm^{-3} $}

\newcommand{\xrayion}{\,$ \rm s^{-1}~H_2^{-1}$}

\newcommand{\hh}{H$_2$}

\usepackage{natbib}
\usepackage{graphics,graphicx}
\usepackage{xcolor}  
\usepackage{natbib}
\usepackage{graphics,graphicx}
\usepackage{xcolor}  

\shorttitle{Time Dependent Water Chemistry}
\shortauthors{Waggoner, Cleeves}

\begin{document}

\title{Modeling Time Dependent Water Chemistry Due to Powerful X-ray Flares from T-Tauri Stars}

\correspondingauthor{Abygail R. Waggoner}
\email{arw6qz@virginia.edu}

\author{Abygail R. Waggoner}
\affil{Department of Chemistry, University of Virginia \\
Charlottesville, VA 22904, USA} 

\author{L. Ilsedore Cleeves}
\affil{Department of Chemistry, University of Virginia \\
Charlottesville, VA 22904, USA} 
\affiliation{Department of Astronomy, University of Virginia \\
Charlottesville, VA 22904, USA}

\begin{abstract}

Young stars emit strong flares of X-ray radiation that penetrate the surface layers of their associated protoplanetary disks. 
It is still an open question as to whether flares create significant changes in disk chemical composition. We present models of the time-evolving chemistry of gas-phase 
 \water\ during X-ray flaring events. The chemistry is modeled at point locations in the disk between 1 and 50\,au at  vertical heights ranging from the mid-plane to the surface.
We find that strong, rare flares, i.e., those that increase the unattenuated X-ray ionization rate by a factor of 100 every few years, can temporarily increase the gas-phase \water\ abundance relative to H can by more than a factor of $\sim3-5$ along the disk surface (Z/R $\ge$ 0.3). We report that a ``typical" flare, i.e., those that increase the unattenuated X-ray ionization rate by a factor of a few every few weeks, will not lead to significant, observable changes.
 Dissociative recombination of \htop, \water\ adsorption and desorption onto dust grains, and ultraviolet photolysis of \water\ and related species are found to be the three dominant processes regulating the gas-phase \water\ abundance. While the changes are found to be significant, we find that the effect on gas phase water abundances throughout the disk is short-lived (days). 
Even though we do not see a substantial increase in long term water (gas and ice) production, the flares' large effects may be detectable as time varying inner disk water `bursts' at radii between 5 and 30\,au with future far infrared observations.
 
 \end{abstract}

\keywords{astrochemistry -- protoplanetary disks -- X-rays: stars  -- molecular processes -- stars: flare} 	

 %			INTRODUCTION
 %%%%%%%%%%%%%%%%%%%%%%%%%%%%%%%%%%%	
%%%%%%%%%%%%%%%%%%%%%%%%%%%%%%%%%%%%
%%%%%%%%%%%%%%%%%%%%%%%%%%%%%%%%%%%% 
 
\section{Introduction}

The composition of planets is likely affected by the physical and chemical processes that occur during their formation in the protoplanetary disk stage. It is therefore necessary to understand these processes in order to create accurate models of planetary system formation.  Observations have shown that protoplanetary disks are flared dust and gas rich disks with radii as large as hundreds of au \citep{ardila2002}. Disks are composed of three layers: 
{\em i.} the photon-dominated region (PDR), which is rich in atomic and ionic species,
{\em ii.} the warm molecular layer, where molecular and radical species can exist \citep{aikawa2002}, and 
{\em iii.} the mid-plane, where molecular and atomic ices can form on dust grains \citep{bergin2007}.
Photons can easily penetrate the surface layers of the disk since it is a low density area with little shielding.  Some UV and X-ray photons can penetrate the warm molecular layer, fostering active chemistry, but few photons reach the dense mid-plane due to the high levels of extinction in this region \citep{bethell2011, fogel2011}. 

The presence of these high energy photons influences the chemistry of the disk, which is thought to evolve slowly over the disk lifetime of $\sim3-10$ Myr \citep{strom1989, glassgold1997, haisch2001, fedele2010}. Chemical species observed in disks are primarily small and simple molecules like CO, HCO$^+$, CN, H$_2$O, OH, CO$_2$, HCN, CS, C$_2$H, and N$_2$H$^+$, and some complex molecules, such as formaldehyde, methanol, and methyl cyanide \citep{dutrey1997, dutrey2007, aikawa2003, oberg2015, walsh2016}. The chemistry of many of these species is directly or indirectly related to the presence or absence of energetic photons or particles from the star and/or environment.

The stellar radiation environment is not constant in time. 
In fact, short-term X-ray flaring is common in young solar mass stars \citep[e.g.,][]{montmerle1996}, and can produce enough X-rays to dynamically affect the chemical and physical properties of the disk \citep{glassgold2005}.
There is some possibility that short-term X-ray activity can impact the chemical composition of the disk. 
\citet{Ilgner2006} considered the theoretical impact of flares on the disk ionization fraction by computing electron abundances using a full chemical model. They found flares could play a significant role in the ionization fraction of the disk, which directly impacted the extent of disk ``dead zones.''
\citet{cleeves2017} found evidence for variations in the disk molecular ion abundances based on multi-epoch submillimeter wavelength data of H$^{13}$CO$^+$ emission in the IM Lup protoplanetary disk. Since this molecule is directly and efficiently formed from H$_3^+$, one possible explanation for the observed time variability is chemistry induced by an X-ray flare.

Among molecules detected in disks, \water\ is one of the most important, owing to its role in the formation of habitable planets like Earth. In addition \water\ ice, referred to here as \waterg, may enable grains to more efficiently stick together when they collide \citep{wang2005}. For planets formed by core accretion, it is believed that both terrestrial and gas giants' rocky inner cores are formed from such
 inelastic grain collisions \citep[e.g.,][]{pollack1996}. Without an ice coating, the grains are more likely to bounce off each other in an elastic collision, which suggests that planet formation can be impacted by the presence of water-rich ice.

 Like \hcop, \water\ is sensitive to the ionization state of gas, as most processes that form water involve the initial ionization of \HH\ \citep{cleeves2014}.
We present a theoretical study of how water chemistry can be impacted by X-ray flares emitted from a T Tauri star, a star similar to the Sun in the early stages of its life. 
We adapt a model based on the physical structure of a disk where variability has previously been detected, IM Lup \citep{cleeves2017},  a solar mass T Tauri star \citep{panic2009}.
The present paper aims to understand how a dynamic X-ray radiation field affects both instantaneous ($\sim$days) and long-term ($\sim$Myr) \water\ abundances in protoplanetary disks, and to identify key species and processes contributing to \water\ abundance changes during energetic flares. 

\begin{table}[t!]
\begin{center}
\caption{Initial Chemical Abundances}\label{initials}
\begin{tabular}{lr|lr}\\\toprule
\hh			&	$5.0 \times 10^{-1}$		&	Grain		&	$6.0 \times 10^{-12}$	\\
\waterg		&	$8.0 \times 10^{-5}$		&	CO			&	$1.3 \times 10^{-4}$		\\
O			&	$1.0 \times 10^{-8}$		&	C			&	$5.0 \times 10^{-9}$		\\
O$_2$		& 	$1.0 \times 10^{-8}$		&	NH$_3$		&	$8.0 \times 10^{-8}$		\\
He			&	$1.4 \times 10^{-1}$		&	HCN			&	$1.0 \times 10^{-8}$		\\
N$_2$		&	$3.75 \times 10^{-5}$	&	C$^+$		&	$1.0 \times 10^{-9}$		\\
CN			&	$6.0 \times 10^{-8}$		&	HCO$^+$		&	$9.0 \times 10^{-9}$		\\
H$_3$$^+$	&	$1.0 \times 10^{-8}$		&	C$_2$H		&	$8.0 \times 10^{-9}$		\\
S$^+$		&	$1.0 \times 10^{-11}$	&	CS			&	$4.0 \times 10^{-9}$		\\
Si$^+$		&	$1.0 \times 10^{-11}$	&	SO			& 	$5.0 \times 10^{-9}$		\\
Mg$^+$		& 	$1.0 \times 10^{-11}$	& 	Fe$^+$		&	$1.0 \times 10^{-11}$	\\	
\hline \noalign {\smallskip}
\hline \noalign {\smallskip}
\end{tabular}
\tablecomments{Abundances are presented with respect to number of 
 hydrogen atoms, and values are a combination of the result of astrochemical modeling of clouds \citep{aikawa99}, with updated protostellar ice abundances.
 }
\end{center}
\end{table} 

 %		MODEL/THEORY
%%%%%%%%%%%%%%%%%%%%%%%%%%%%%%%%%%%%
%%%%%%%%%%%%%%%%%%%%%%%%%%%%%%%%%%%%
%%%%%%%%%%%%%%%%%%%%%%%%%%%%%%%%%%%%
\section{Model}\label{theory}

\subsection{Disk Model}\label{diskmodel}

To model the chemistry in a dynamic X-ray environment we adopt the model from \citet{fogel2011}, updated in \citet{cleeves2014}, which includes 647 species and 5944 reactions and processes. 
The code adopts the rate equation method 
and calculates the non-equilibrium chemistry as a function of time. 
\begin{table}[t]
\begin{center}
\caption{\water\ Related Processes}\label{H2Oreactions}
\begin{tabular}{rcll}\\\toprule
%%%%%%%%%%%%%%%%%%%%%%%%%%%%%%%%%%%%%%%%%%%
 \multicolumn{4}{c}{\textbf{Dissociative Recombination}}							\\
 %%%%%%%%%%%%%%%%%%%%%%%%%%%%%%%%%%%%%%%%%%%	
{1.}	& {	\htop\ + $e^-$} 					& {\arw} 	& {\water\ + H}			\\	
{2.}	& {	\htop\ + Grain$^-$}				& {\arw}	& {\water\ + H + Grain	}	\\	
%%%%%%%%%%%%%%%%%%%%%%%%%%%%%%%%%%%%%%%%%%%
\hline \noalign {\smallskip} 
 \multicolumn{4}{c}{\textbf{Photo-Chemistry}}		 							\\
%%%%%%%%%%%%%%%%%%%%%%%%%%%%%%%%%%%%%%%%%%%
{3.}	& {	\water\ + $\gamma_{UV}$}		& {\arw}	&  {\waterp\ + $e^-$}	\\
{4.}	&{	\water\  + $\gamma_{UV}$}		& {\arw}	& {OH + H}				\\
{5.}	&{	\water\  + $\gamma_{UV,fl}$}		& {\arw}	& {OH + H}				\\	
{6.} 	&{	\water\  + $\gamma_{UV,fl}$}		& {\arw}	& {O + \HH}				\\	
%%%%%%%%%%%%%%%%%%%%%%%%%%%%%%%%%%%%%%%%%%%
\hline \noalign {\smallskip} 
 \multicolumn{4}{c}{\textbf{Neutral + Neutral} }									\\
%%%%%%%%%%%%%%%%%%%%%%%%%%%%%%%%%%%%%%%%%%%
{7.}	& {	H + OH}						& {\arw}	& {\water\ }				\\	
{8.}	&{	\HH + OH} 					& {\arw}	& {\water\ + H}			\\
{9.}	&{	OH + OH	}					& {\arw}	& {\water\ + O}			\\
{10.}	&{	H + \water\ }					& {\arw	}	& {OH + \HH}			\\
%%%%%%%%%%%%%%%%%%%%%%%%%%%%%%%%%%%%%%%%%%%
\hline \noalign {\smallskip} 
 \multicolumn{4}{c}{\textbf{Ion + Neutral} }										\\

%%%%%%%%%%%%%%%%%%%%%%%%%%%%%%%%%%%%%%%%%%%
{11.}	&{	O$^-$ + \HH}					& {\arw}	& {\water\ + $e^-$}		\\
{12.}	&{	OH$^-$ + H }					& {\arw} 	& {\water\ + $e^-$}		\\
{13.}	&{	H$^+$ + \water\	}			& {\arw	}	& {\waterp\ + H	}		\\
{14.}	& {	C$^+$ + \water\	}			& {\arw}	& {HOC$^+$ + H	}		\\ 
{15.}	&{	C$^+$ + \water\	}			& {\arw	}	& {HCO$^+$ + H	}		\\
{16.}	&{	HCO$^+$ + \water\	}			& {\arw}	& {\htop\ + CO	}		\\
%%%%%%%%%%%%%%%%%%%%%%%%%%%%%%%%%%%%%%%%%%%
\hline \noalign {\smallskip} 
 \multicolumn{4}{c}{\textbf{Adsorption and Desorption} }						\\
%%%%%%%%%%%%%%%%%%%%%%%%%%%%%%%%%%%%%%%%%%%
{17.}	&{	\waterg\ + Heat	}				& {\arw}	& {\water\	}			\\
{18.}	& {	\waterg\ + $\gamma_{Ly\alpha}$}	& {\arw}	& {\water\	}			\\
{19.}	& {	\waterg\ + $\gamma_{UV}$}		& {\arw}	& {\water\	}			\\ 
{20.}	&{	\water\ + Grain	}				& {\arw	}	&{ \waterg\  + Grain	}	\\
\hline \noalign {\smallskip} 
\hline \noalign {\smallskip} 
\end{tabular}
\tablecomments{
Twenty processes contributing to the production and consumption of gas-phase \water\ in the layers of the disk exposed to X-ray flares. 
Note the absence of grain surface water formation among the top reactions governing water formation and destruction. Grain surface chemistry is expected to be more important closer to the mid-plane and radially further out than in the high radiation regions of the disk focused on in the present work.
 }
\end{center}
\end{table} 
The code begins with the initial chemical abundances motivated by interstellar cloud models (Table \ref{initials}), which are designed to be representative of molecular cloud abundances. The code runs for 0.5\,Myr to reach a pseudo steady state equilibrium at the time of the flare, where 0.5\,Myr is redefined as $t=0$. 
The code then slows down to calculate the chemistry during the event with fine resolution (30 minute time steps). 
All chemical abundances are presented with respect to total number of hydrogen atoms.

Physical parameters of the IM Lup protoplanetary disk were used in this model, the first source to show significant chemical variability \citep{cleeves2017}. 
The model inputs include gas density ($\rho$), gas temperature (T$_{\rm gas}$), dust temperature (T$_{\rm dust}$), UV flux, and X-ray ionization rate, and are sampled from the \citet{cleeves2016} IM Lup disk model.
The simulations were run at discrete points labeled by their radial distance from the star ($R$) and vertical height from the mid-plane ($Z/R$), see Appendix \ref{conditions}. 
The disk is assumed to be azimuthally symmetric and reflected about the mid-plane.
The model treats the point locations independently and does not take into account interactions between horizontal zones. 

Parameters for the central star are as follows: effective temperature of T$_{\rm eff}=3900$ K, stellar radius of 2.5 R$_\odot$ \citep{pinte2008}, stellar mass of 
1 M$_\odot$ \citep{panic2009}, which is held fixed as was done in \citet{cleeves2016}. 

The \citet{smith2004} Ohio State University (OSU) gas-phase chemical network was used as a basis for the chemical network, and the method of \citet{hasegawa1992} was used to model grain-surface chemistry. The six types of chemical and physical processes most relevant for the present paper include:
\begin{enumerate}
\item dissociative recombination reactions, where a molecular ion accepts a negative charge and dissociates (Table \ref{H2Oreactions}, Reactions 1 and 2),
\item photolysis, where high energy photons dissociate  or ionize species (Table \ref{H2Oreactions}, Reactions 3-6),
\item neutral + neutral association reactions, defined as \newline $A+B$ \arw\ $ C+D$ (Table \ref{H2Oreactions}, Reactions 7-10),
\item neutral + ion association reactions, defined as \newline $A + B^{+/-}$ \arw\ $C + D^{+/-}$ (Table \ref{H2Oreactions}, Reactions 11-16),
\item gas-grain adsorption and desorption
(Table \ref{H2Oreactions}, Processes 17-20),
\item grain surface chemistry, such as $A_{(gr)} + B_{(gr)}$ \arw\ $ AB_{(gr)} $ (not shown in Table \ref{H2Oreactions}).
\end{enumerate}

Four types of photons were included in the model for photolysis reactions. 
UV Photons (\uvphoton) produced by the star's accretion shock \citep{gullbring1998}; 
Ly$\alpha$ photons ($\gamma_{\mbox{Ly}\alpha}$), which can carry up to 85\% of the UV flux \citep{herczeg2004}; 
X-ray photons produced by the central star; 
UV photons (\xrayphoton) produced when electrons are ejected during X-ray ionization of H$_2$, that impact H$_2$ and create a fluorescent UV field that can dissociate and ionize chemical species \citep{maloney1996}.

\begin{figure*}[th!] 
\begin{center}
\includegraphics[scale=0.84]{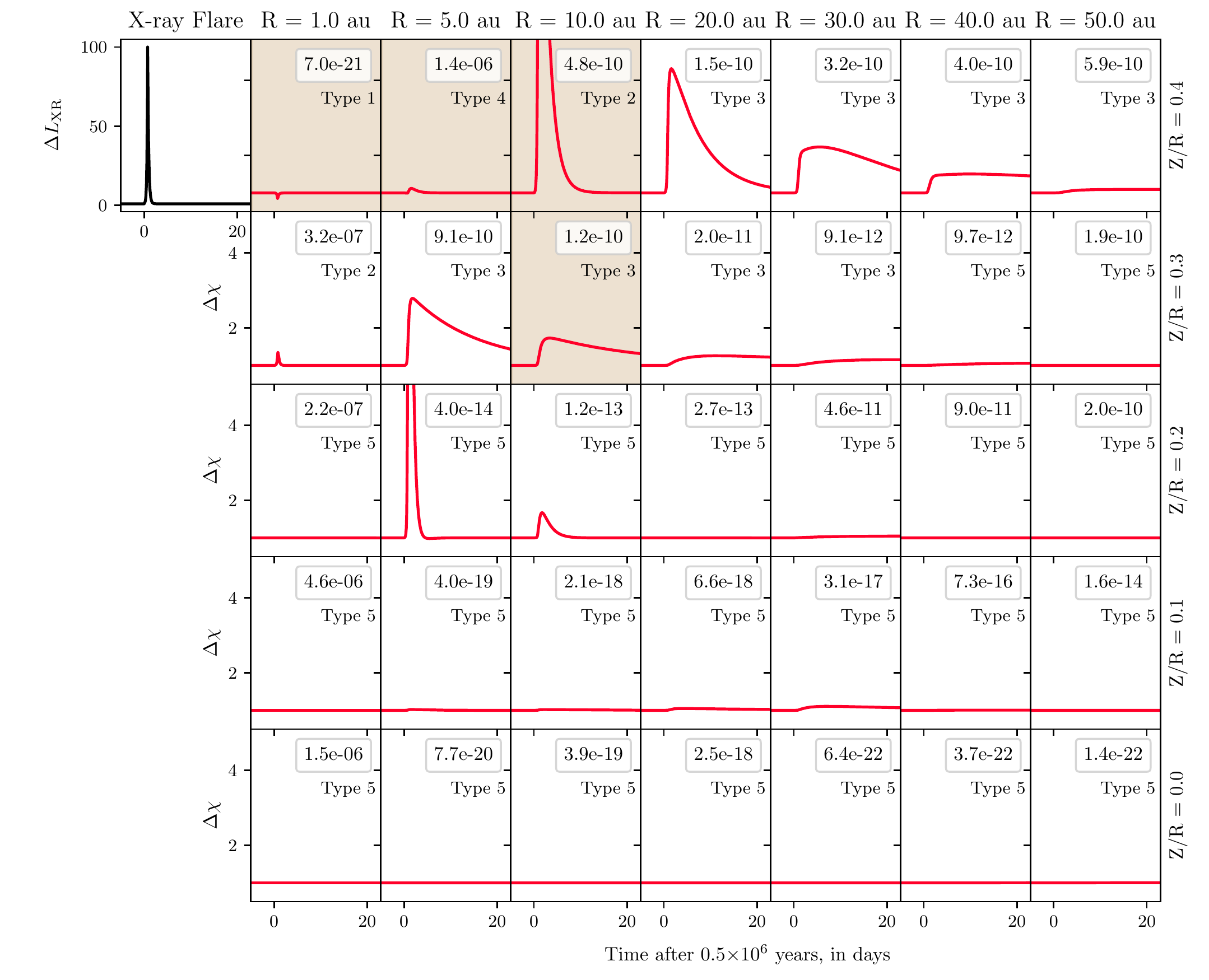} 
\end{center}
\caption{
\water\ response grid from Test 1. See \S \ref{methods} for test parameters. Initial \water\ abundances appear at the top of each plot and correspond to $\Delta \chi = 1$. Curve types are described in \S \ref{responses}, and represent the five different response curves seen in the model.}
For the points where the maximum abundance is not visible on the plot, the maxima occur: 1. $\rm (R, Z/R) = (10\,au,0.4)$ at $t = 1.3$ days, $\Delta \chi = 13.4,$ 2. $(R,Z/R) = (5\,au, 0.2)$ at $t = 1.1$ days, $\Delta \chi = 15.5$. 
Analyzed points representing Curves 1-4 are highlighted in tan. Detailed analysis is presented in \S \ref{responses}.
\end{figure*}\label{T1grid}

%					METHODS
%%%%%%%%%%%%%%%%%%%%%%%%%%%%%%%%%%%%
%%%%%%%%%%%%%%%%%%%%%%%%%%%%%%%%%%%% 
%%%%%%%%%%%%%%%%%%%%%%%%%%%%%%%%%%%%
\subsection{Chemical Analysis During Flares}\label{methods}
The chemical response to X-ray flares was examined at a total of 35 points with radii of $R=1, 5, 10, 20, 30, 40$, and $50$\,au and normalized vertical heights of $Z/R = 0.0, 0.1, 0.2, 0.3$, and $0.4$ (Appendix \ref{conditions}, Figure \ref{dentem}). These points were chosen since they are are relatively close to the star, and the vertical heights allow us to examine chemical responses in the different regions of the disk.
We additionally explored models beyond 50\,au and saw relatively little change as discussed below in Section \S \ref{results}.
The cosmic ray rate was held constant at a low value of 2.0$\times$10$^{-20}$ ionizations per H$_2$ per second. However, \citet{cleeves2013} determined that cosmic rays are energetically negligible compared to X-rays at the disk surface, the region focused on in this paper, and so the specific value of this input parameter does not impact our findings.
 See Table \ref{points} for physical values at all twenty points.
Changes in chemical abundance per H atom are determined by plotting $\Delta$$\chi$, 
which represents the relative change in abundance compared to a model with no X-ray flares, versus time and is defined as:
\[\Delta \chi = {{\Delta \chi}_{\rm{with \ flare}} \over {\Delta \chi}_{\rm{without \ flare}}}\] 
 It should be noted that for a given flare, highly abundant ($\geq 10^{-5}$) molecules do not show relative change as strongly as those with low abundances (e.g., $\lesssim 10^{-20}$ per H). 
Individually, these models show \water\ responses at specific locations in the disk, but together they create a 2D model mapping out \water\ flare responses in radius and height, as shown in Figure \ref{T1grid}. 

 The peak X-ray flare strength used in our fiducial model is 100 times the magnitude of the baseline X-ray luminosity, which was $L_{\rm XR} = 4\times10^{30}$~erg~s$^{-1}$. This high flare value was chosen to show the maximum impact on water abundance due to X-ray flares. The flare rise time is 3 hours, and the exponential decay time is 5 hours. These timescales are chosen to be typical of observed X-ray flares from T-Tauri stars \citep[e.g.,][]{pe2005}.
 We additionally ran models for both single flare and multi-flare events.
 Test 1 was run with a single flare initiated at 0.5\,Myr ($t=0$\,days).
We define the beginning of the flare as six e-folding times prior to the flare peak, effectively setting the flare strength equal to zero at $t=0$.
Test 2 was run with two flares in rapid succession at 0 days and 5 days.
 Each test was modeled for approximately 23 days after $t=0$.
 Note that the timescales presented in \S \ref{results} do not include light travel time, so it assumed that the X-ray ionization rate at each time step reaches each location in the disk instantaneously.

%					RESULTS
%%%%%%%%%%%%%%%%%%%%%%%%%%%%%%%%%%%%
%%%%%%%%%%%%%%%%%%%%%%%%%%%%%%%%%%%%
%%%%%%%%%%%%%%%%%%%%%%%%%%%%%%%%%%%%

\section{Results}\label{results}

To guide the discussion in the following sections, Table \ref{H2Oreactions}  lists the key reactions and processes at the twenty points of the disk considered here. The following sections will refer back to these processes. 

\subsection{General Trends Throughout the Disk}\label{ptrends} 

The model revealed that \water\ responses beyond 50\,au are insignificant compared to distances interior to 50\,au.
 At distances greater than 50\,au there are X-ray flare responses, but responses are typically $\leq$ 10\%, and the \water\ abundance is $\leq 10^{-10}$ at most points. As such, we focus the remainder of our analysis on radii $\le 50$ au.

Figure \ref{T1grid} shows \water\ responses at 35 points across the disk inside of 50\,au. Several trends are seen in the responses both vertically and radially. 
The mid-plane is along $Z/R=0$, and has no flare response. This behavior is expected because photons are absorbed in the PDR and warm molecular layer before reaching the mid-plane. 
Likewise, response strength decreases radially from the inner disk atmosphere to the outer disk atmosphere as X-rays become geometrically diluted with distance from the star.

 The effect of the snow line, the point at which \water\ exists primarily in the solid rather than gas-phase 
  state, can also be seen as drops in \water\ abundances across the disk. 
Generally, any points where gas-phase
  \water\ abundance is less than $10^{-5}$ are exterior to the snow line, where \waterg\ is the dominant phase of \water. 
  The surface, such as the point $(R,Z/R) = (1.0\,\rm au, 0.4)$  is an exception, since this region is strongly impacted by UV radiation from the star and neither gas-phase
 nor ice-phase  \water\ exist in high abundance due to the high levels of photodissociation. 

\subsection{Analysis of Different \water\ Response Curves}\label{responses}

Five types of time-varying abundance curves of \water\ are observed in the disk. All 35 positions are labeled with their closest response curve type in Figure \ref{T1grid}, alongside the adopted X-ray light curve. The responses include:
\begin{enumerate}
\item abundance decrease, such as at $(R,Z/R) = (1.0\,\rm au, 0.4)$, where the response curve is a narrow negative spike, 
\item production with a fast exponential rise and decay ($t < 10$\,days), such as $(R,Z/R)$ = $(10.0\,\rm au, 0.4)$, where the response curve is a positive sharp rise,
\item initial production with a slow decay ($t > 10$\,days), such as $(R,Z/R)$ = $(10.0\,\rm au, 0.3)$, where the curve increases and experiences little to no decay,  
\item hybrid, such as $(R,Z/R)$ = $(5.0\,\rm au, 0.4)$, where the curve decreases to a minimum below the initial abundance then increases to a maximum spike, 
\item and no observable flare response, such as $(R,Z/R)$ = $(1.0\, \rm au, 0.0)$, where $\Delta \chi = 0$ or $\Delta \chi \leq \pm0.05$ and abundance $< 10^{-13}$ with respect to H atom. 
\end{enumerate}
These curve types are further analyzed in the following sections. Moreover, the seven fastest processes involved in the production and destruction of gas-phase \water\ were analyzed for each response type using Table \ref{H2Oreactions} for reference.

%	TYPE 1: SHORT TERM DESTRUCTION
%%%%%%%%%%%%%%%%%%%%%%%%%%%%%%%%%%%%

\subsubsection{Type 1: Short-term Destruction}\label{type1}   

Only one point in the disk shows a net destruction curve, $(R,Z/R) = (1.0\,\rm au, 0.4)$, shown in
 Figure \ref{typeonef}. 
 Water abundance drops by 15\% after 0.75 days, then quickly increases to approximately its initial abundance. 
 
Detailed rates for reactions and processes involving \water\ are evaluated at three representative times, before the flare ($t=0 $ days), at the peak \water\ response
($t=0.75$ days), and after the peak \water\ response ($t=3 $ days). 
Of all the disk points modeled, this one is the closest to the star and at the disk surface. This point is more strongly irradiated  than other points in the disk, and the increase in X-ray photons during the flare tends to enhance the \water\ photodissociation rate. This behavior can be seen by the appearance of 
Reactions 5 and 6 
at $t=0.75$ days. 
These two reactions involve \water\ destruction by UV photons that arise from X-ray photon induced \HH\ fluorescence (\xrayphoton). 
At $t=0$ days and $t=3$ days Reactions 5 and 6 are not among the seven fastest processes and only appear during the time of the flare.
After the flare ends \water\ is reformed by Reactions 7 and 8 to return to its initial abundance.

The abundance of OH was expected to increase since it is a product of Reactions 4 and 5: however, Figure \ref{typeonef} 
shows that OH abundance decreases. There are two explanations as to why. First, Reactions 4 and 5 are not among the seven fastest 
OH reactions, so it is likely that these are not the dominant OH production mechanisms.
In addition, OH experiences multiple destructive photolysis reactions, including OH ionization and dissociation to form O and H.
\begin{figure}
\begin{center}
\includegraphics[scale=0.56]{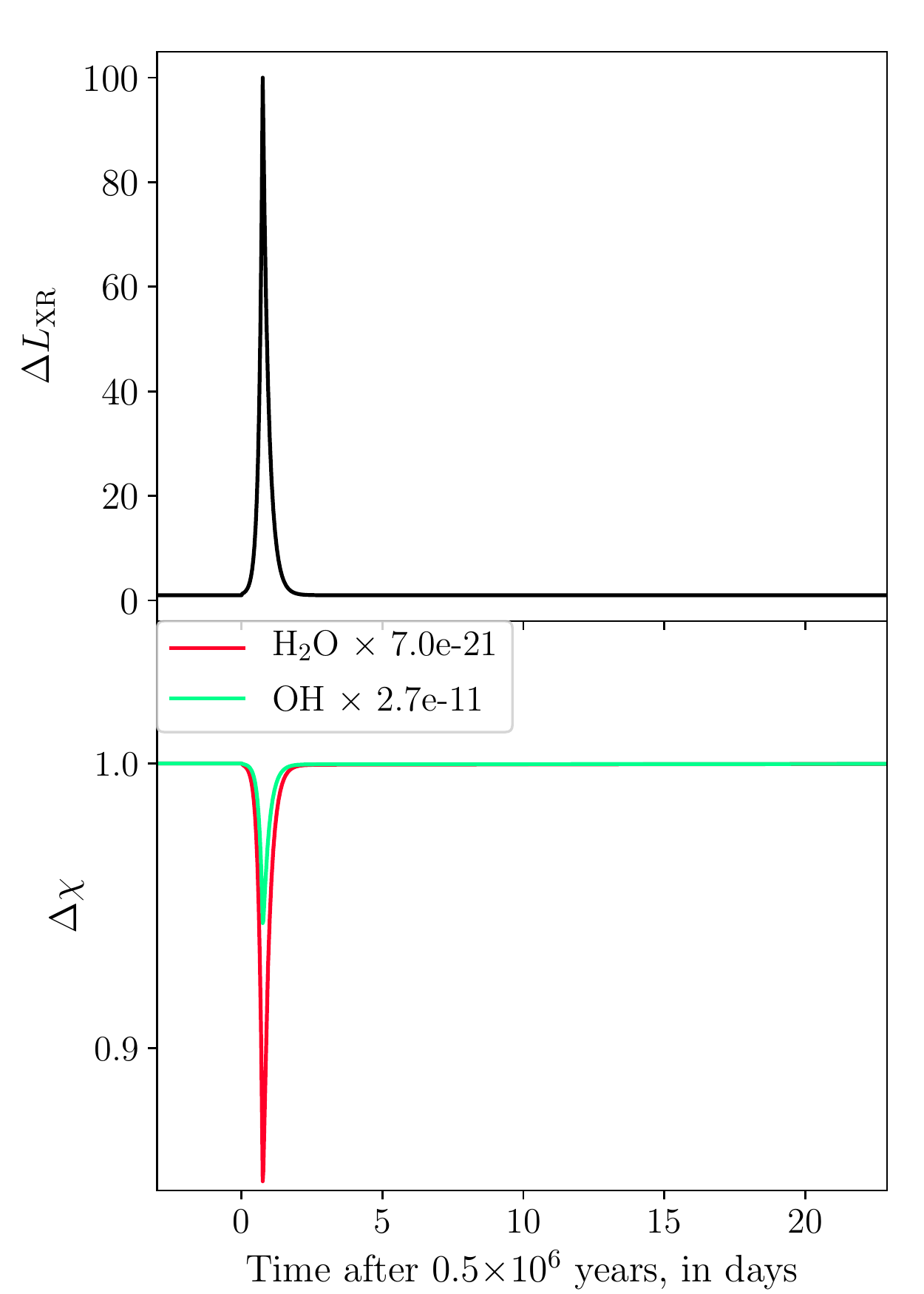}
\end{center}
\caption{Type 1 Reaction Curve: Destruction, $(R,Z/R) = (1.0\,\rm au, 0.4)$. \water\ is temporarily destroyed through photolysis to produce OH (Reaction 4), but \water\ is reformed by Reactions 7 and 8 to return to its initial abundance. OH abundance decreases due to rapid photolysis.}
\end{figure}\label{typeonef} 
 \begin{figure}
\begin{center}
\includegraphics[scale=0.56]{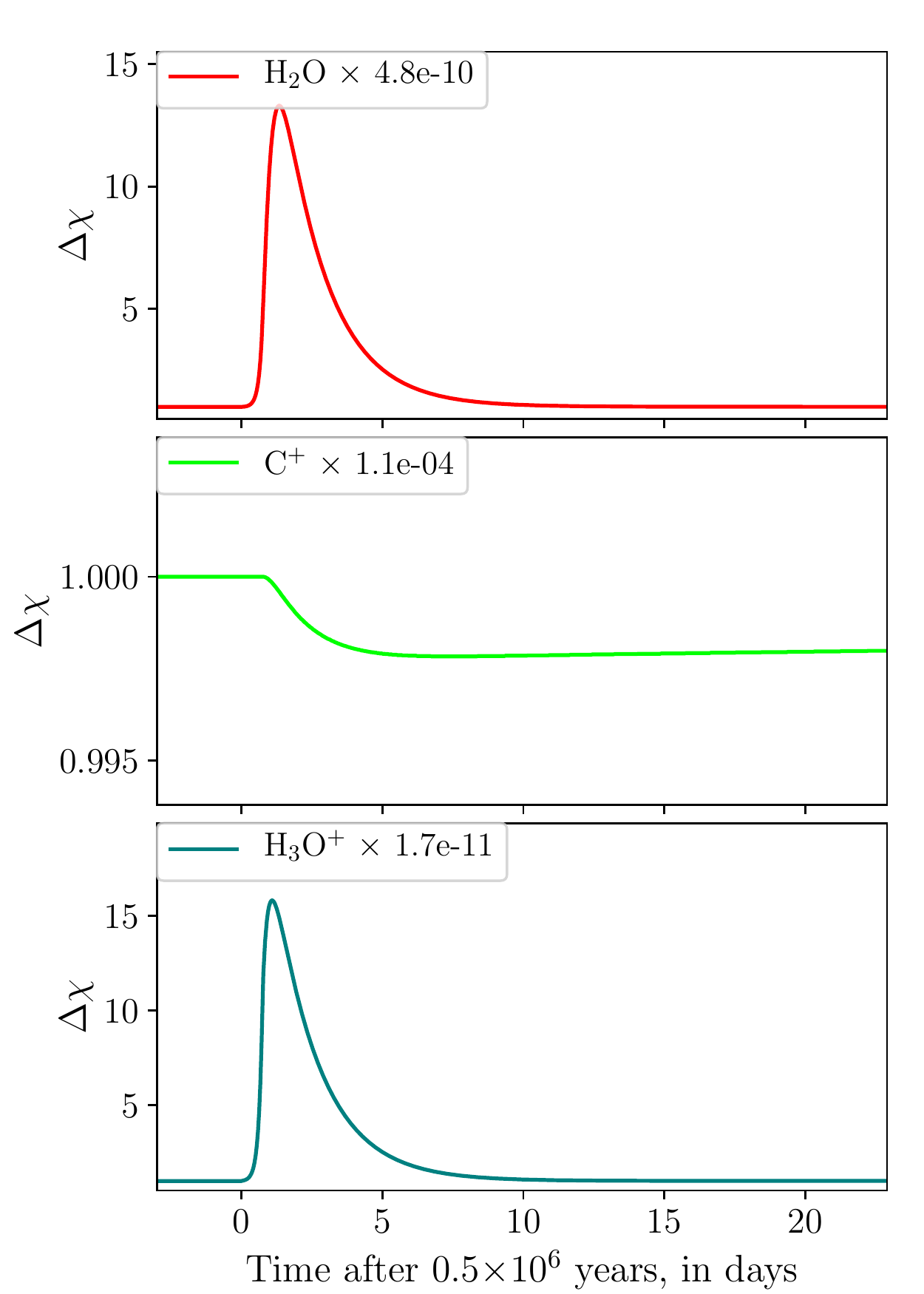}
\end{center}
\caption{Type 2: Production with fast rise and decay ($<10$\,days), $(R,Z/R) = (10.0\,\rm au, 0.4)$.
 Dominant \water\ production is dissociative recombination of \htop\ (Process 1).
 Dominant \water\ consumption is ion-neutral reaction with C$^+$. Refer to Figure \ref{typeonef} for change in X-ray ionization rate. 
}
\end{figure}\label{typetwof}

%TYPE 2: RAPID PRODUCTION PRODUCTION WITH FAST DECAY
%%%%%%%%%%%%%%%%%%%%%%%%%%%%%%%%%%%% 

\subsubsection{Type 2: Rapid Production with Fast Decay ($<10$\,days)}\label{type2}

Type 2 responses, i.e., a rapid production with fast rise and decay, occur commonly in the disk grid (Figure \ref{T1grid}).
 The strongest response is at $(R,Z/R) = (10.0\,\rm au, 0.4)$, shown in Figure \ref{typetwof}. 
At this point \water\ abundance increases $\sim 1340\%$ at 1.3 days, 
then returns back to its initial abundance after 10 days. 

Detailed rates are evaluated at four representative times, before the flare ($t=0 $ days), at the peak \water\ response ($t=1.25 $ days and $t=4.35$ days), and after the peak \water\ response ($t=8 $ days). 
This point is both on the surface layer of the disk and radially close enough to feel 
a strong impact from the X-ray flare, but not so strong that the flare is solely destructive like in Type 1. Because of this, neutral and ionic atoms and some molecules, such as C$^+$ and \htop\ are abundant enough to be significant in chemical reactions. 

Figure \ref{typetwof} 
reveals that the \htop\ abundance increases by over 1500\% in direct response to the flare, which then speeds up the dissociative recombination of \htop, which in turn drastically increases the production of \water\ (Reactions 1 and 2).
There is a flare response delay from \water\ because it takes time for \htop\ to find electrons to dissociatively recombine to form \water. Reaction 8 also contributes to the production of \water\ due to an increase of 1230\% of OH due to the flare. Reaction 8 is an order of magnitude slower than dissociative recombination, and thus does not impact the increase of \water\ abundance as strongly as Reaction 1. 

Once the X-ray flare ends, \htop\ returns back to its initial abundance, Reaction 1 rate significantly drops at $t=4.35$ days, which decreases \water\ production. 
Consumption is much faster than production 
since \htop\ is no 
longer overly abundant to produce \water. The main destructive process is an ion-neutral reaction between \water\ and C$^+$ (Reactions 14 and 15), with some contribution from UV photolysis (Process 4). C$^+$ is produced directly from the UV flux, thus \water\ is primarily consumed from photolysis and photolysis dependent processes at this location. 
Once \water\ abundance has returned to its initial abundance the pseudo steady 
state equilibrium is reached and further changes in abundance do not occur. 

%	TYPE 3: Longer Lived Enhancement
%%%%%%%%%%%%%%%%%%%%%%%%%%%%%%%%%%%% 

\subsubsection{Type 3: ``Long-lived'' Enhancement ($ > 10$\,days)}\label{type3} 

Response curves that experience an initial enhancement with a slow decay over the course of the simulation ($> 10 $ days) are observed along the 
$Z/R = 0.4$ and $Z/R = 0.3$ vertical heights (Figure \ref{T1grid}). The point $(R,Z/R) = (10.0\,\rm au, 0.3)$ was chosen to represent this trend, and whose evolution in plotted in more detail in Figure \ref{type3f}. 
After a 0.2 day delay, \water\ increases 73\% at 3.3 days, then steadily decreases until the end of the simulation. 
Detailed rates are evaluated at four representative times, before the flare ($t=0 $ days), after the flare and before/at the peak \water\ response ($t=1.25 $ days/$t=4.35$ days), and after the peak \water\ response ($t=8 $ days) 

\begin{figure}
\begin{center}
\includegraphics[scale=0.57]{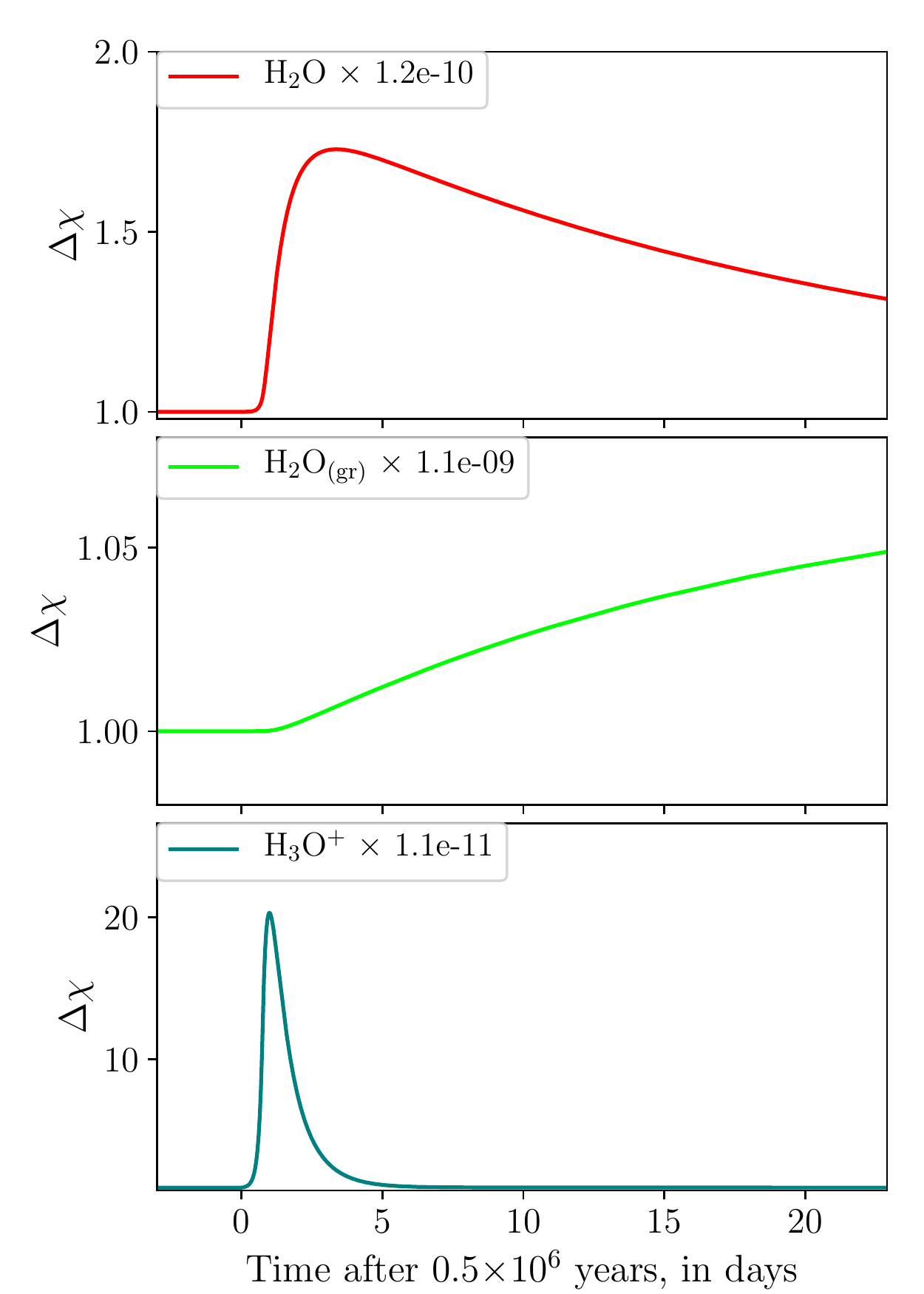}
\caption{Type 3: Long-lived enhancement ($> 10$ days) $(R,Z/R) = (10.0\,\rm au, 0.3)$. Main \water\ production is dissociative recombination of \htop\ (Reactions 1 and 2). Main \water\ consumption is adsorption (Process 20). Refer to Figure \ref{typeonef} for change in X-ray ionization rate. 
}
\end{center}
\end{figure}\label{type3f}

Like a Type 2 response (\S \ref{type2}), the initial increase in \water\ abundance is caused by a sharp increase in \htop\ 
abundance (Figure \ref{type3f} shows that \htop\ abundance increases over 2000\%).
 \htop\ undergoes dissociative recombination via Reactions 1 and 2 to produce \water\ faster than \water\ is consumed.
While photolysis does occur, this point is not as strongly exposed to the destructive photons, so the fastest gas-phase removal process of excess gas-phase \water\ is adsorption onto grains, creating a slower tail of \water\ removal from the gas. 
\begin{figure}[th!]
\begin{center}
\includegraphics[scale=0.56]{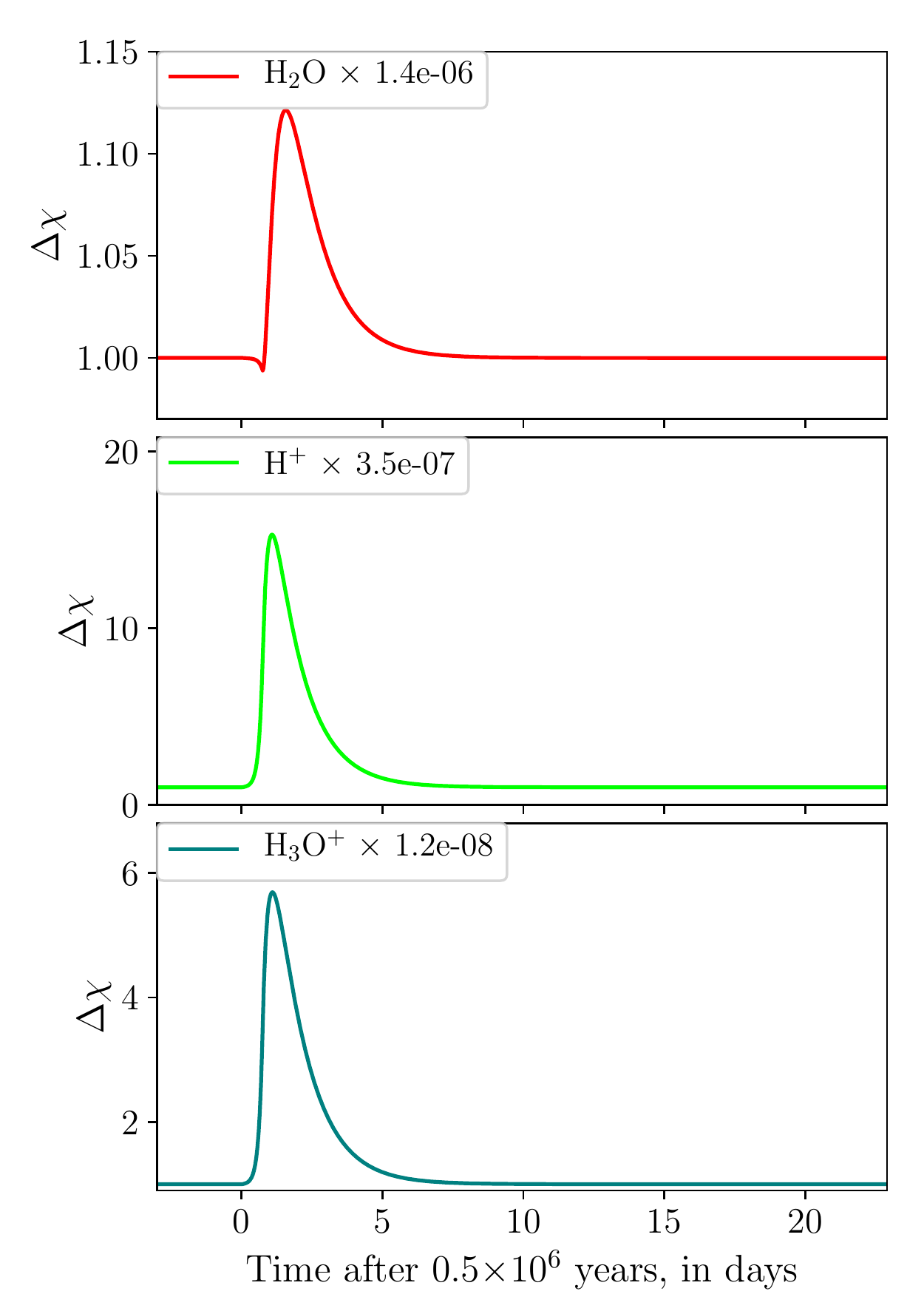} 
\caption{
Type 4: Hybrid, $(R,Z/R) = (5.0\,\rm au, 0.4)$. 
Initially, \water\ is consumed by an increase in photolysis reactions (Processes 5 and 6), then is produced by dissociative recombination of \htop\ (Process 1), then it is consumed by a combination of photolysis (Processes 4, 5, and 6) and ionization (Process 13) to return to its initial abundance. Refer to Figure \ref{typeonef} for change in X-ray ionization rate. 
}
\end{center}
\end{figure}\label{typefourf}
Figure \ref{type3f} shows that \waterg\ is steadily increasing (to $\sim$4\%) as \water\ is steadily decreasing. It is to be noted that \water\ decline appears greater than \waterg\ increase for two reasons. 
First, \waterg\ is more abundant than \water, so the change in abundance is smaller,
 and second, \water\ is also being consumed via multiple processes including slow UV destruction, so not all gas-phase
  \water\ is being converted to ice.  Longer simulations are required to determine if the water ice abundance remains permanently elevated or returns back to its initial abundance.

%					TYPE 4: HYBRID
%%%%%%%%%%%%%%%%%%%%%%%%%%%%%%%%%%%% 

\subsubsection{Type 4: Hybrid}\label{type4}

Unlike the other curve types, the hybridized response experiences both an absolute minimum and an absolution maximum. $(R,Z/R) = (5.0\,\rm au, 0.4)$ is the only hybrid response observed in this model (Figure \ref{typefourf}). 
\water\ abundance initially decreases $\sim$1\%
 at $t=0.5$ days, then increases $\sim$12\%
  above initial abundance at $t=1.6$ days, then returns back to its initial abundance for the duration of the run (15 days).
Detailed rates are evaluated at four representative times, before the flare ($t=0 $ days), at the absolute minimum \water\ response, ($t=0.5 $ days), at the absolute maximum \water\ response ($t=1.7$ days), and after the peak \water\ responses ($t=5 $ days). 

 Similar to the Type 1 response at $(R, Z/R) = (1.0\, \rm au, 0.4) \,(\S \ref{type1})$, this point is strongly impacted by the flare since it is at a close distance to the star (5.0\,au) and on the disk surface.
 However, it is far enough away from the star to have a high abundance of gas-phase molecules and ions, such as \water\ and \htop ($\geq 10^{-8}$ per H).
  Because of its proximity to the star and high abundance of gas-phase precursors, the response of \water\ is a combination of Type 1 (\S \ref{type1}) and Type 2 (\S \ref{type2}) curves. 
During the initial flare impact, \water\ abundance decreases due to an increase photodissociation (Processes 5 and 6), similar to Type 1. 
 After the initial impact, \water\ is then produced via the normal gas-phase channels followed by dissociative recombination of \htop, similar to Type 2. 
  An initial decrease in \water\ abundance is only observed at $(R,Z/R) = (5.0\,\rm au, 0.4)$ due to the initial enhanced photodissociation by X-ray induced UV fluorescence. 
At the end of the flare \water\ is no longer being produced by Process 1 as quickly, since \htop\ abundance decreases (Figure \ref{typefourf}). \water\ then returns to its initial abundance due to a combination of photolysis and charge exchange by \Hp\ (Process 13) followed by dissociative recombination of \waterp.

\begin{center} 
\begin{figure}
\includegraphics[scale=0.59]{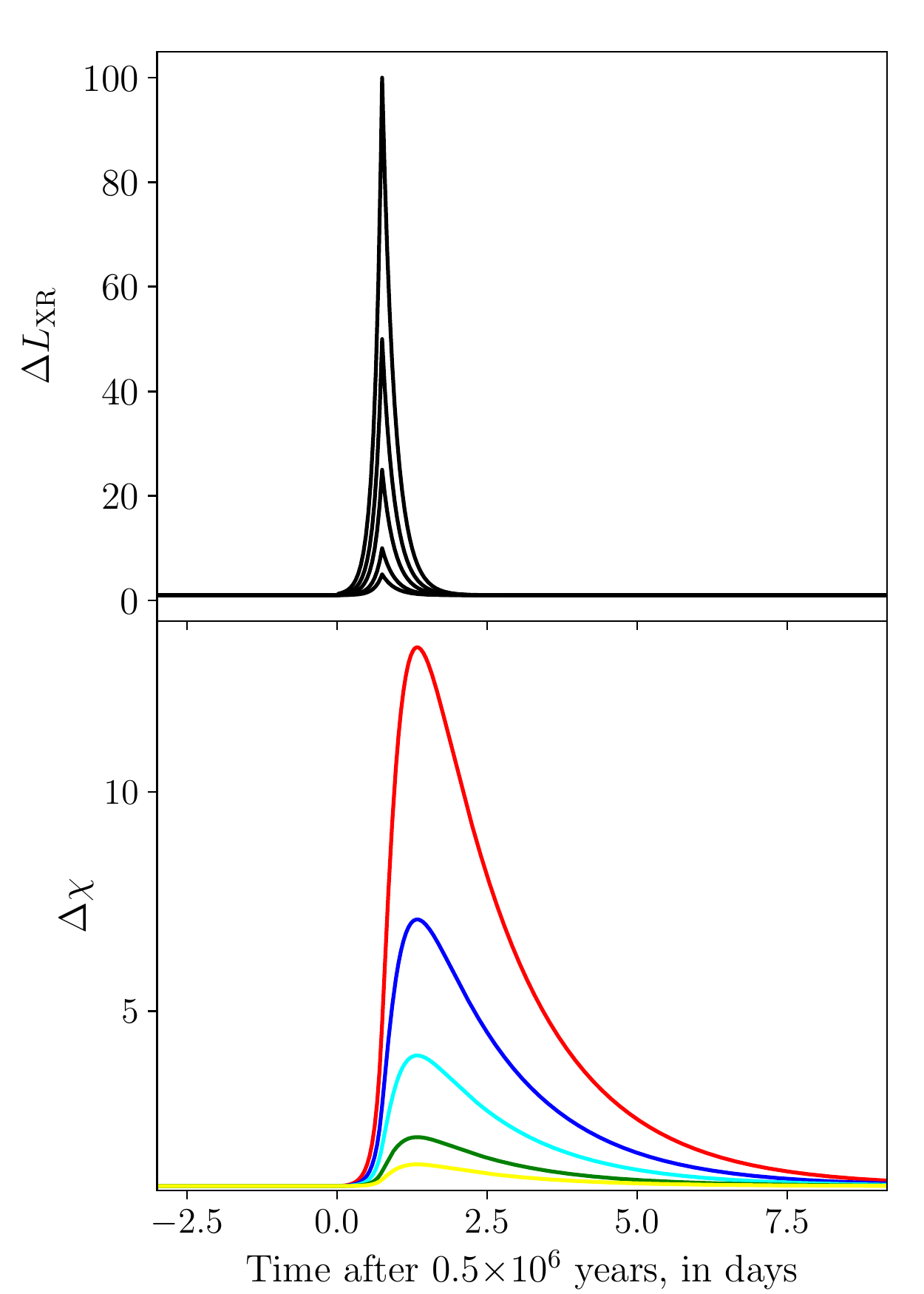} 
\caption{\water\ responses to flares of different peak magnitudes at R = 10au Z/R = 0.4. Red indicates flare strength 100, blue indicates 50, cyan indicates 25, green indicates 10, and yellow indicates 5 times the background X-ray ionization rate. Note that the overall abundance profile stays similar while the peak \water\ abundance changes with respect to the X-ray flare strength.}
\end{figure}\label{104overlap}
\end{center}
\begin{figure}
\begin{center}
\includegraphics[scale=0.68]{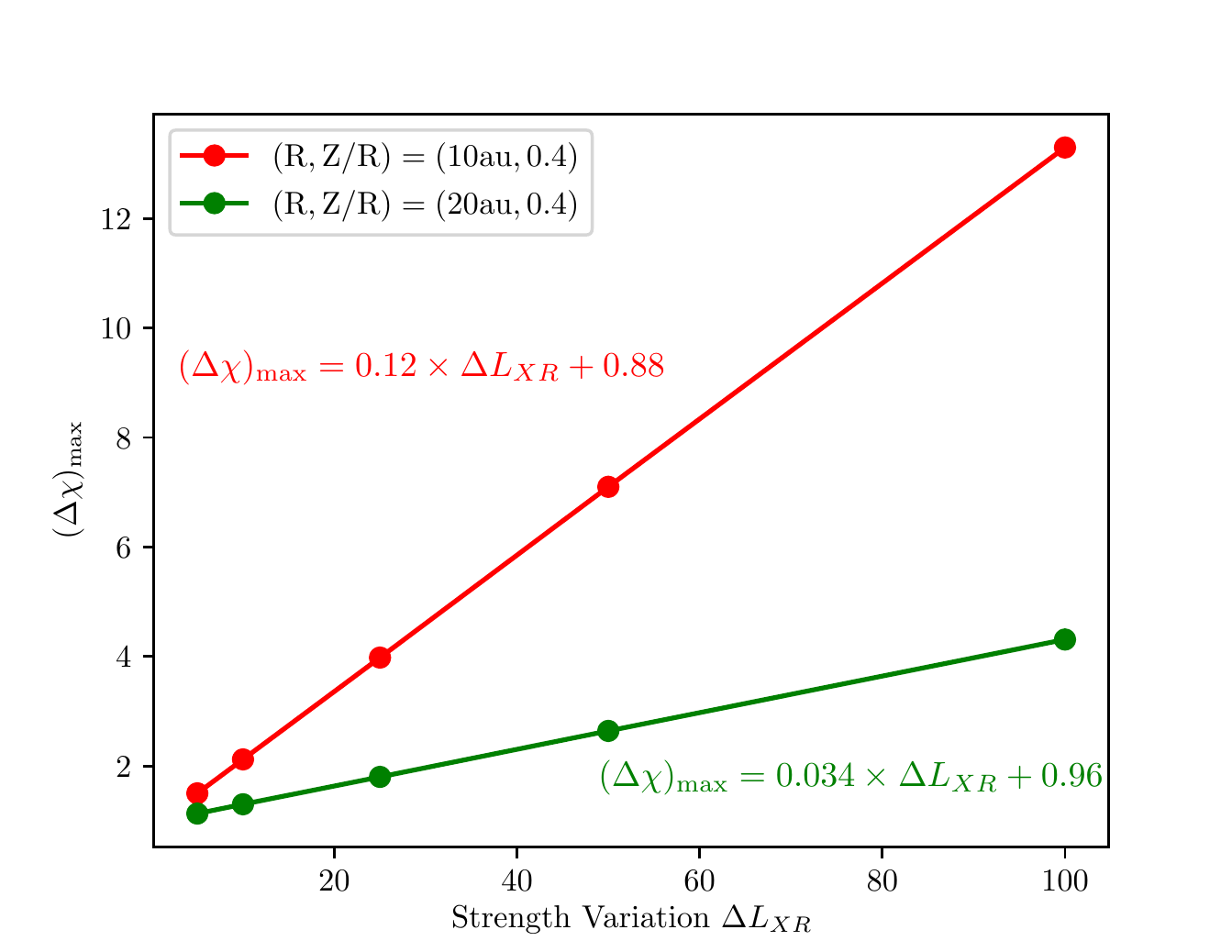} 
\end{center} 
\caption{
The maximum change in \water\ abundance that occurs due to peak flare strengths of 100, 50, 25, 10, and 5 times the relative background X-ray ionization rate}, along with linear fits as indicated.
\end{figure}\label{variation_plot}

%			TYPE 5: NO RESPONSE
%%%%%%%%%%%%%%%%%%%%%%%%%%%%%%%%%%%%

\subsubsection{Type 5: No Observable Response}\label{type5}

Observable flare responses do not occur at the most dense regions of the disk ($\rho \geq 10^{-12}$\density) or along the mid-plane for one of two reasons. In regions where there is a large abundance of gas-phase
 \water\ ($\geq 10^{-7}$ per H) \water\ is not as sensitive to change. In regions where there is a low X-ray ionization rate (essentially zero) there are not enough photons present to initiate chemistry.

\begin{figure*}[th!] 
\begin{center}
\includegraphics[scale=0.83]{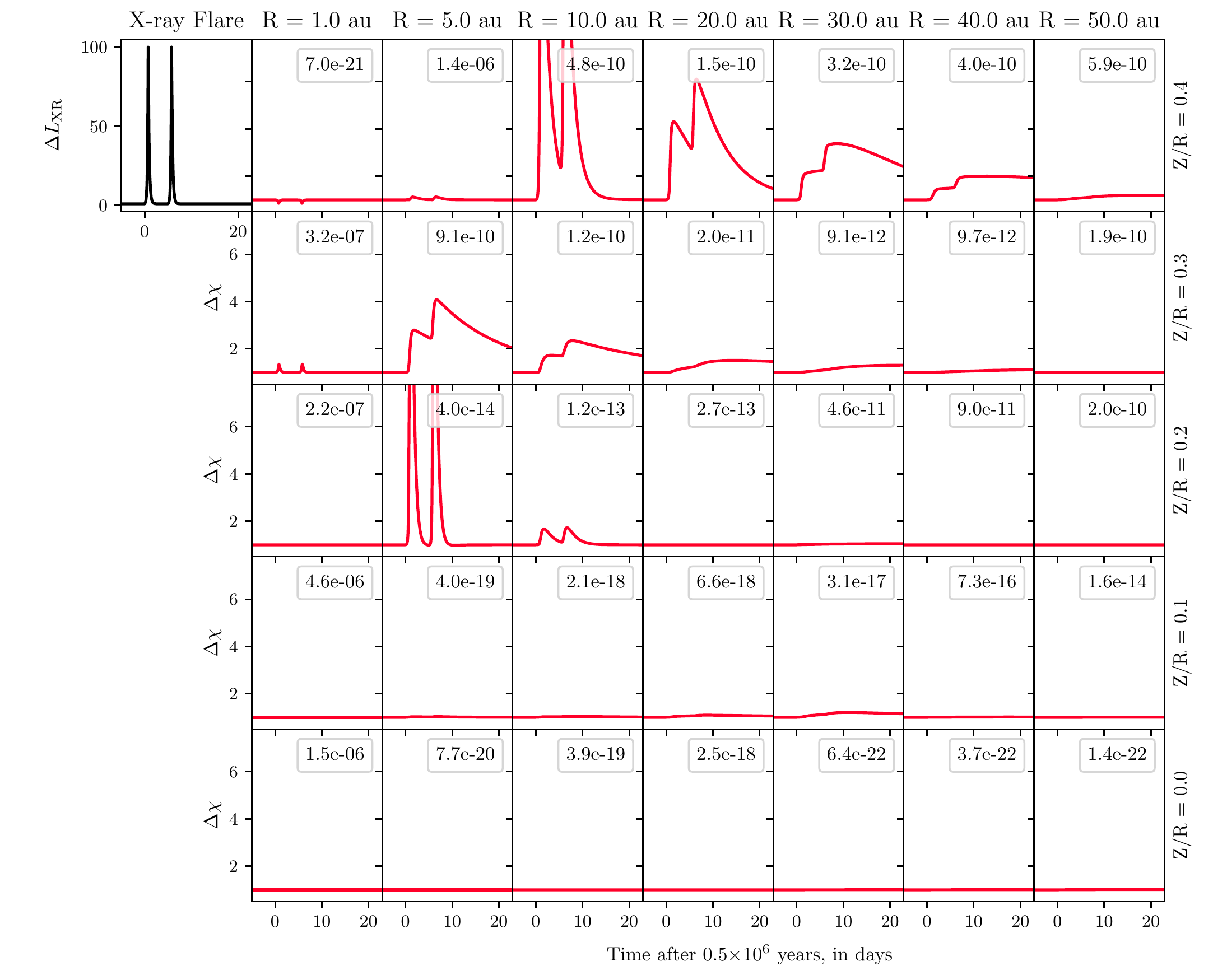}
\end{center}
\caption{\water\ response grid from Test 2. See \S\ref{methods} for test parameters.} Flares occur at $t=0$\,days and $t=5$\,days. Maximums at $(R, Z/R) = (10, 0.4)$ occurs at $\Delta \chi = 12.6, t = 1.7$ days and $\Delta \chi = 13.4, t = 6.6$ days. Second maximum at $(R, Z/R) = (20, 0.4)$ occurs at $\Delta \chi = 5.8, t = 7.5$ days.
\end{figure*}\label{T2grid}

% 		FLARE STRENGTH VARIATION
%%%%%%%%%%%%%%%%%%%%%%%%%%%%%%%%%%%
\subsection{Flare Strength Variation}\label{strength}

All previous analyses were performed with X-ray flare strengths equal to 100 times the background X-ray ionization rate, which is considered an extremely strong and rare flaring event \citep{wolk2005, getman2008}.
We found that weaker flares (a few times X-ray ionization rate), which are much more common, produced similar time-varying abundance profiles in this model, but with smaller amplitude. 

Figure \ref{104overlap} shows \water\ responses to flare strengths of 5, 10, 25, 50 and 100 at $(R,Z/R)$ = $(10\, \rm au, 0.4)$. Figure \ref{variation_plot} presents a relationship between the maximum change in \water\ abundance that occurs and flare strength, for values of 100, 50, 25, 10, and 5 times the baseline X-ray luminosity. 
Both $(R, Z/R) = (10\, \rm au, 0.4)$ and $(20\, \rm au, 0.4)$ were found to have a linear dependence on change in \water\ abundance compared to the change in flare strength (Figure~\ref{variation_plot}).
The maximum change in the \water\ abundance in response to a flare follows the following relationships: at
 $(R, Z/R) = (10\, \rm au, 0.4)$,  $\Delta\chi=0.12 \times\Delta L_{XR} +0.88$, and at   $(R, Z/R) = (20\, \rm au, 0.4)$,  $\Delta \chi =0.034 \times \Delta L_{XR}+0.97$,
 where $\Delta L_{XR}$ is the peak change in X-ray ionization rate.

%		MULTI-FLARES
%%%%%%%%%%%%%%%%%%%%%%%%%%%%%%%%%%%

\subsection{Impacts of Muliflare Events}\label{flares}

In addition to the single flare analysis (Test 1, Figure \ref{T1grid}), we investigate a model that has multiple
 flaring events (Test 2, Figure \ref{T2grid}) discussed qualitatively here.
Between Tests 1 and 2 only slight variations occurred. 
As seen by comparing Figures \ref{T1grid}  and \ref{T2grid} the plots have similar response patterns. Figure \ref{T2grid} 
shows that a response to two flares behaves in a similar pattern to a response to one flare, only the responses ``stack up" on each other. This behavior suggests that multiple flares will have a similar effect as a single flare, and their cumulative effect will be related to the timing between flares.

\begin{figure*}[th!] 
\begin{center}
\includegraphics[scale=0.77]{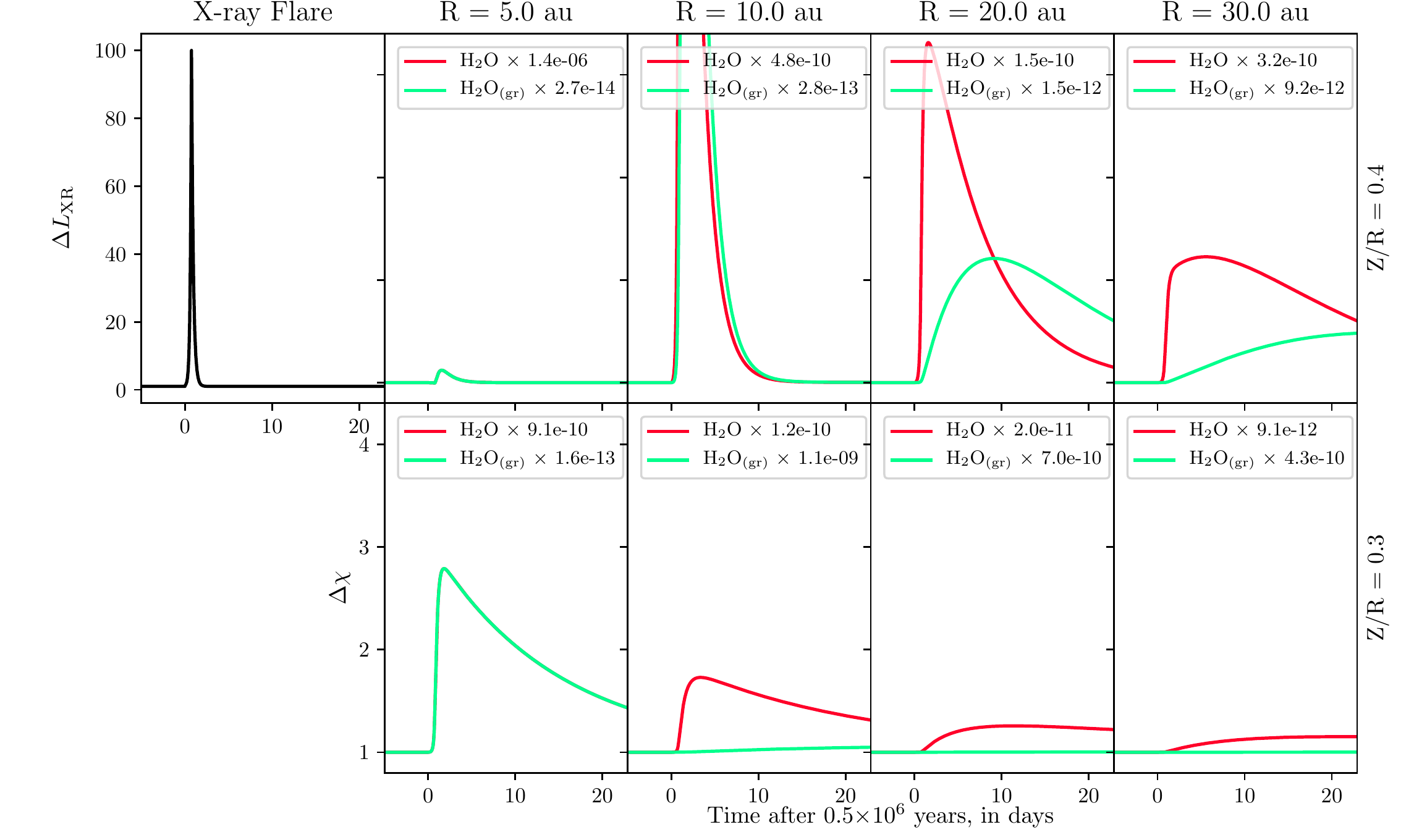} 
\caption{Gas-phase
 \water\ and \waterg\ responses radially along the vertical heights of $Z/R = 0.4$ and $0.3$, between $R=5$\,au and $30$\,au. 
 For the point $\rm (R, Z/R) = (10\,au,0.4)$, where the maximum abundance is not visible on the plot, the \waterg\ maximum occurs  at $t = 2.3$ days and $\Delta \chi =10.8$ and the gas-phase \water\ maximum occurs at $t = 1.3$ days, $\Delta \chi = 13.4$.
}
\end{center}
\end{figure*}\label{H2O(gr)}

%			ICE WATER RESPONSES
%%%%%%%%%%%%%%%%%%%%%%%%%%%%%%%%%%%%

\subsection{\water\ Ice Responses}\label{gvss}

In the following section, we quantify how much of the gas-phase response is imparted into the ice.
\waterg\ responses discussed in this section are obtained from the fiducial strong X-ray flare model. The results of the runs for \waterg\ are shown in Figure \ref{H2O(gr)}.

Interior to the snow line there is little \waterg, and the small amount that exists has the same flare response as the gas. When the dust temperature decreases to below the desorption temperature of \water, $\lesssim 170\,$K, \waterg\ responses start to deviate from gas-phase \water. Finally, in colder regions ($\sim 20 - 100\,$K), ice and gas-phase responses show little to no correlation.

 This suggests that as excess \water\ is produced in hot regions by flares, \water\ rapidly adsorbs onto grains to form \waterg. As the excess gas-phase \water\ returns back to its initial pre-flare abundance, \waterg\ desorbs and does so as well (Processes 17-19, Table \ref{H2Oreactions}). At temperatures just below \water\ freeze-out ($\lesssim 170\,$K), desorption is a much slower process, resulting in the observed ``longer-lived" response in \waterg\ compared to gas-phase \water. 
 
 The cold, shielded regions, including the points from 5 to 30 au at $Z/R = 0.3$, there is a small but persistent amount \waterg\ for the duration of the simulation. At $(R, Z/R) = (30.0\,\rm au, 0.4)$ the increase is much more significant, about 50\%, and does not decrease within the duration of the simulation.
 
 We note that changes to \waterg\ from discrete flaring events do not strongly impact the total amount of \waterg\ the disk, since the majority of \waterg\ lies near the midplane and long term surface changes are minimal (up to 3\%) but may have cumulative effects that will be explored in future work.

%			DISCUSSION
%%%%%%%%%%%%%%%%%%%%%%%%%%%%%%%%%%%%
%%%%%%%%%%%%%%%%%%%%%%%%%%%%%%%%%%%%
%%%%%%%%%%%%%%%%%%%%%%%%%%%%%%%%%%%%

\section{Discussion}\label{discussion}

%		RESPONSE TRENDS
%%%%%%%%%%%%%%%%%%%%%%%%%%%%%%%%%%%%

\subsection{Response Trends in the Disk}\label{rtrends}
 
General reaction trends across the disk can be found by comparing the vertical and radial response trends in the disk with the five curve types discussed in the previous Section (\S \ref{responses}) and the physical conditions of the gas as summarized in Table \ref{points}.
A Type 1 response is rare and only occurs in regions along the surface of the disk and close ($\sim$1\,au) to the central star, where the X-ray ionization rate is highest.

Along the vertical height traced by $Z/R=0.3$, response curves behave primarily with a radially fading Type 3 to Type 5 (no response) pattern.
The region closest to the star at $(R,Z/R)=(\rm 1.0\,au, 3)$ is an exception to this trend, since it has a much greater \water\ abundance (three orders of magnitude) and is located in a region with a high UV flux. 

A second trend occurs along the $Z/R=0.3$ vertical height: as radial distance increases, the response curves begin to develop a more prominent lagging tail (Type 3). This feature is likely because photon intensity decreases as distance from the star increases, so after the initial \water\ production, the weaker radiation field decreases the rate of photodissociation. Thus, the dominant gas-phase \water\ consumption process transitions to adsorption onto grains. These Type 3 responses show that \water\ abundance can increase for tens of days, or more. However, additional longer simulations show that each Type 3 response does eventually return to its initial abundance within a year. 

Along the vertical heights traced by $Z/R=0,$ and $1$ no significant response is observed. This height is along and near the mid-plane, and it is very difficult for photons to penetrate since they are absorbed in the upper layers, so the X-ray ionization rate is very low ($\leq 10^{-27}$\xrayion).

%		POTENTIAL OBSERVABILITY  
%%%%%%%%%%%%%%%%%%%%%%%%%%%%%%%%%%%%

\subsection{Potential Observability}\label{observability}

The ideal conditions to observe X-ray driven variability on \water\ requires both \water\ to be sufficiently abundant ($>10^{-10}$ per H) and a sufficiently large response in magnitude, at the very least 5\% or greater.  In addition, having a moderate-long lasting effect can help confirm an abundance change. If the change in abundance occurs for too short of a time, one would have to be very lucky to catch it. Alternatively, if the changes are too gradual, they may go unnoticed. 
 
The regions that show the greatest magnitude of change in water abundance are those with X-ray ionization rates $\geq 10^{-17}$\xrayion\ and densities of $\leq 10^{-12}$\density. When X-ray ionization rates exceed $\geq 10^{-10}$\xrayion\ Type 1 (destructive) response curves are observed. 
Correspondingly, the significantly changed regions that are potentially observable lie near the surface of the disk ($Z/R \geq 0.3$) and between 10\,au and 40\,au distance from the star. 
These regions can be observed using the ground state transitions of water in the far infrared at $557$ GHz ($1_{11}-0_{00}$) and $1113$ GHz ($1_{10}-1_{01}$).

 \begin{figure}
\begin{center}
\includegraphics[scale=0.7]{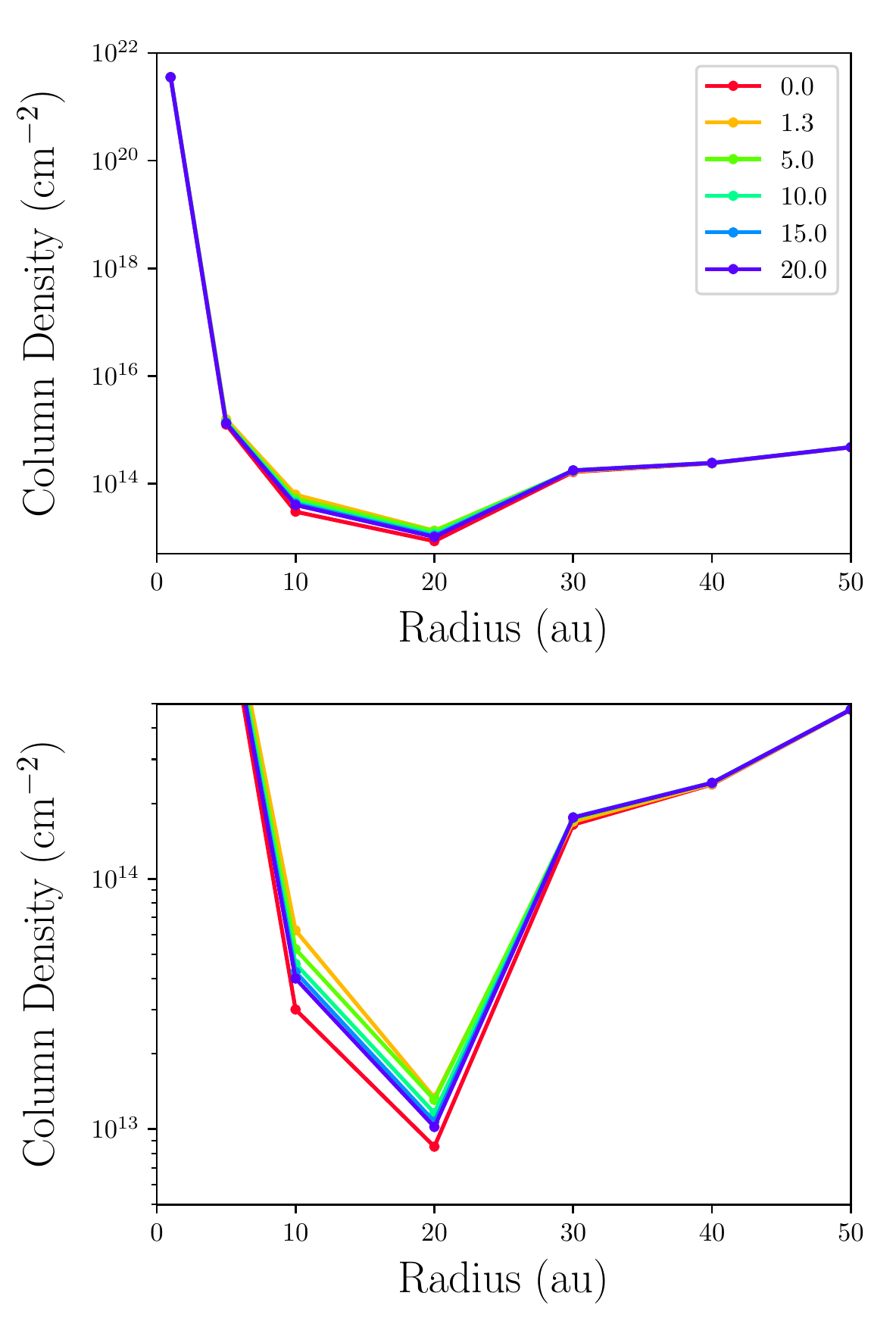}%plotted using col_den_plot.py
\caption{
\water\ column densities, N(\water), are shown at times before flare impact (t=0.0 days), at maximum N(\water) increase (t=1.3 days), and as N(\water) returns to pseudo steady-state. The right panel shows a zoom-in of the maximally changed region of the left panel, highlighting the factor of $\sim2$ change in column density.}
\end{center}
\end{figure}\label{colden}

Figure \ref{colden} shows the vertical column density of gas phase water at different times after our fiducial strong flare model.
Changes in radial column density presented in Figure \ref{colden} incorporate light travel time. The times are plotted such that $t=0$ is defined as six e-folding times prior to the peak flare at the star's location.
The high column density of water close to the star comes from the water snow line. The intermediate dip and radial rise from 20 to 30\,au is caused by increased photo-destruction of water at these intermediate radii. Between 5 and 30\,au the greatest change in overall column density of gas-phase \water\ is seen, varying by just over a factor of two in column density. Note that even though the change in \water\ abundance at individual heights can be much greater, but the higher densities in the deeper layers, $Z/R=0.3$ and below, more strongly impacts the overall change in column density.

 Within this radial range, \water\ responses have potential observability up to 10 days after flare impact,
  when there is still a 53\% and 36\% increase in N(\water) at $R=10$\,au and $20$\,au. 
\water\ ice variability is unlikely to be observed with little to no variation in column density. 

\subsubsection{Choice of Disk Model}

It is to be noted that IM Lup is extremely massive compared to a ``typical" disk, with 0.17 $M_{\odot}$  \citep{cleeves2016}. A typical disk has a mass of $\sim0.04\, M_{\odot}$ \citep{williams2011}. A lower gas mass results in higher photon penetration due to less absorption along the disk surface. Therefore, more of the disk volume will be exposed to X-ray flares. 

IM Lup also lacks strong disk substructure, with only weak spiral arms and rings in the millimeter emission \citep{huang2018} and rings in the infrared scattered light \citep{avenhaus2018}. We surmise that if such features are associated with low disk surface density (such as gaps), they will allow for further X-ray penetration into the disk. The resulting chemical responses will be more intense and extend closer to the mid-plane than the responses modeled in IM Lup. 
In addition, IM Lup is vertically flared, which intercepts more stellar radiation. X-ray flares may not impact a flat disk as strongly. 

% 			SUMMARY
%%%%%%%%%%%%%%%%%%%%%%%%%%%%%%%%%%%%  
%%%%%%%%%%%%%%%%%%%%%%%%%%%%%%%%%%%%	
%%%%%%%%%%%%%%%%%%%%%%%%%%%%%%%%%%%%

\section{Summary}\label{summary}

We have conducted a theoretical study of protoplanetary disk water chemistry exposed to strong X-ray flaring events. Our models show that the gas-phase  \water\ abundance at the surface responds to X-ray flares in many ways, but the effects are typically short term, lasting a few days or weeks.  Analyzing the chemical reaction rates in detail reveals that \water\ production occurs dominantly by dissociative recombination of \htop, and gas-phase \water\ destruction occurs dominantly by photolysis, adsorption onto grains, or ion-neutral reactions with C$^+$. Photolysis is dominant in the warmer ($T_g>200$~K) upper layers of the disk, whereas adsorption is  dominant in cooler ($T_d<170$ K) deeper layers of the disk. Reactions with C$^+$ were only observed in an ion-rich region along the disk surface. 

Overall the dominant production mechanisms for water formation during X-ray flares are the gas-phase channels. Grain surface chemistry is included in the model, but it appears to be too slow of a process to impact the short term \water\ abundance changes.

The magnitude of the response is dependent on the local X-ray ionization rate and gas density.
 Regions with low gas density have high X-ray ionization rates, which tend to have large and flare responses, while regions with higher density and low X-ray ionization rates tend to have no observable flare response. Between these extremes, we find the largest response on water abundances to flares.

We emphasize, this work focuses on strong X-ray flares to understand the maximum effect of these phenomena on \water. However, these are rare events and occur every few years. We find that more common flares, i.e., those that increase X-ray luminosity by factors of a few, do not significantly impact \water\ in the disk.

It is possible that any observations done during major flaring events would lead to a higher observed abundance of \water\ than would typically be in the present disk. This type of time-dependent chemical variability has already been observed and reported by \citet{cleeves2017} with H$^{13}$CO$^+$ in the IM Lup protoplanetary disk. 

Short-term chemical variability could occur in a range of astronomical objects, other than protoplanetary disks, that are exposed to X-ray flaring. 
Recently, \citet{mackey2019}
modeled non-equilibrium chemistry in molecular clouds caused by X-ray flaring from active galactic nuclei. While conditions used in this model differ greatly from protoplanetary disk environments, it lends further support to the idea that astronomical environments are far more chemically dynamic than previously assumed. 
 
It should also be noted that these simulations are focused on discrete flaring events. Additional work is needed to examine how the full stochastic behavior of flaring from a young star might accumulate water over time, thereby influencing the proclivity for a given disk to form potentially habitable planets.

% 	ACKNOWLEDGEMENTS AND BIBLIOGRAPHY
%%%%%%%%%%%%%%%%%%%%%%%%%%%%%%%%%%%%  
%%%%%%%%%%%%%%%%%%%%%%%%%%%%%%%%%%%%	
%%%%%%%%%%%%%%%%%%%%%%%%%%%%%%%%%%%%
 
 \acknowledgments

The authors are grateful for support from the 2017 Smithsonian Astrophysical Observatory REU program funded by the National Science Foundation REU and Department of Defense ASSURE programs under NSF Grant AST-1659473, and by the Smithsonian Institution.
This material is based upon work supported by the National Science Foundation Graduate
Research Fellowship Program under Grant No. 1842490. Any opinions,
findings, and conclusions or recommendations expressed in this material are those of the
author(s) and do not necessarily reflect the views of the National Science Foundation.

% 		APPENDIX 
%%%%%%%%%%%%%%%%%%%%%%%%%%%%%%%%%%%%  
%%%%%%%%%%%%%%%%%%%%%%%%%%%%%%%%%%%%	
%%%%%%%%%%%%%%%%%%%%%%%%%%%%%%%%%%%%

\appendix

% 	Protoplanetary Disk Physical Conditions
%%%%%%%%%%%%%%%%%%%%%%%%%%%%%%%%%%%% 

\section{Protoplanetary Disk Physical Conditions}\label{conditions}

Within a protoplanetary disk, gas and dust density is highest along the mid-plane and near the central star, and density decreases as the disk surface is approached and radial distance from the star increases as a natural consequence of the viscous motion of the gas \citep{lyndenbell1974}. Temperature, X-ray ionization rate, and UV flux  are highest along the disk surface and decrease with increasing radial distance from the star. Figure \ref{dentem} plots these physical conditions in the IM Lup protoplanetary disk, , which are the physical conditions used in the present study. Points whose chemistry is modeled are represented by red dots in Figure \ref{dentem}. The exact physical condition values used for all 35 modeled points are presented in Table \ref{points}.

\begin{figure}[th!]
\begin{center}
\includegraphics[scale=0.54]{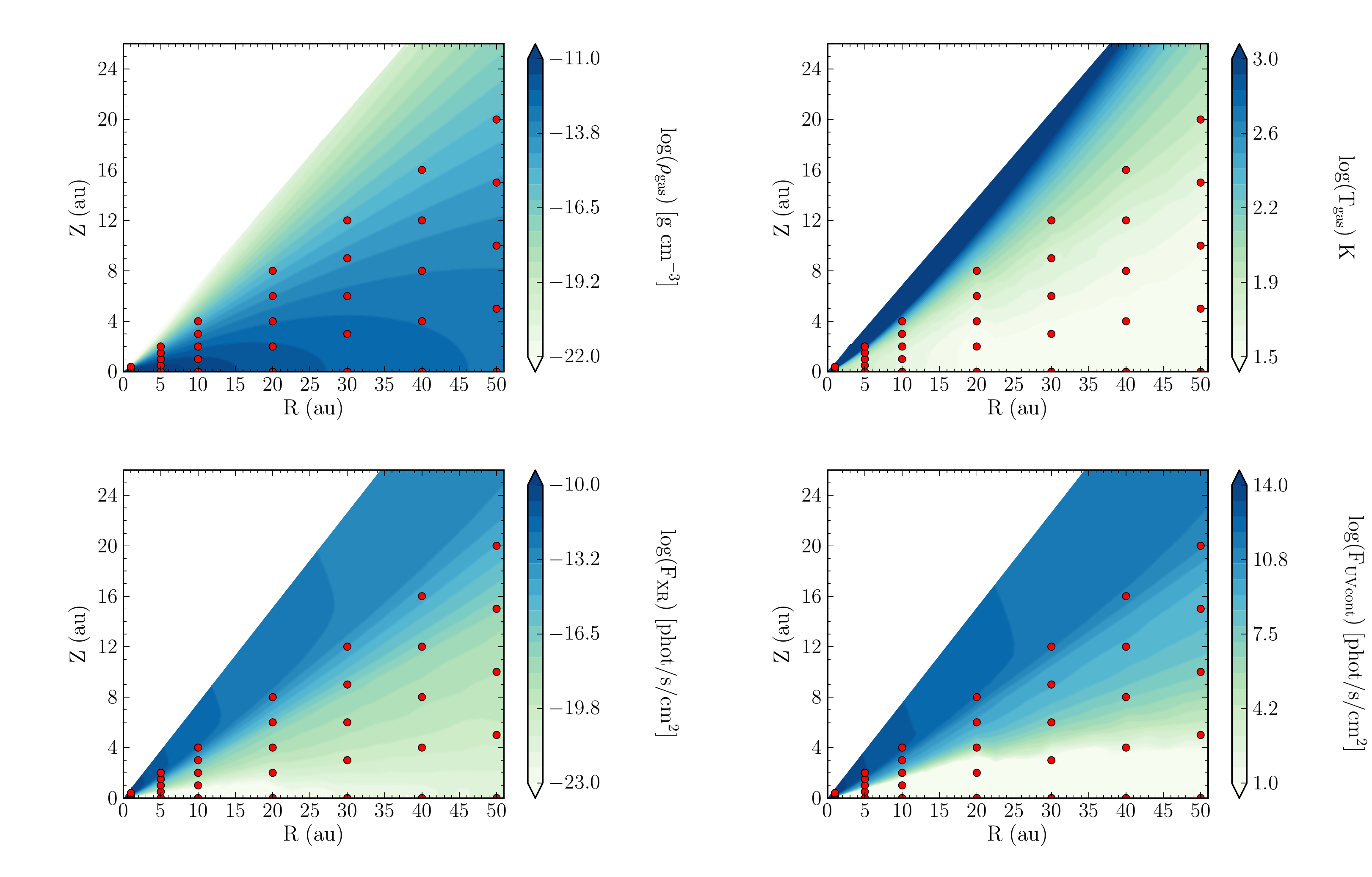}
\end{center}
\caption{These plots represent density, temperature, X-ray ionization rate, and UV flux in the IM Lup protoplanetary disk, a disk encircling a solar-mass young star. Red dots indicate the 35 locations modeled and analyzed in this paper. }
\end{figure}\label{dentem} 
 
\begin{table}[tbh] 
\begin{center}
\caption{Model locations and parameters.}
\begin{tabular}{c|c|c|c|c|c|ccccc}
R		& Z		& $\rho$				& T$_{\rm gas}$& T$_{\rm dust}$	& UV Flux						& X-ray Ionization       			\\
au		& au		& $\rm g cm^{-3} $		& K			& K				& $\rm \gamma s^{-1} cm^{-2}	$	& $ \rm s^{-1} H_2^{-1}$			\\
\hline \noalign {\smallskip}
1.0		& 0.0		& 1.07$\times10^{ -9}$	& 134.3		& 134.3			& 0.0							& 1.0$\times10^{ -30}$			\\
1.0		& 0.1		&  2.49$\times10^{ -10}$	& 132.2		& 133.2			& 0.0							& 1.06$\times10^{ -30}$			\\
1.0		& 0.2		&  7.03$\times10^{ -12}$	& 168.4		& 167.8			& 3.37$\times10^{10}$			& 5.64$\times10^{ -16}$			\\
1.0		& 0.3		&  7.29$\times10^{ -15}$	& 662		& 342.5			& 4.50$\times10^{14}$			& 3.19$\times10^{ -11}$			\\
1.0		& 0.4		& 1.98$\times10^{ -18}$	& 4200		& 341.1			& 5.82$\times10^{14}$			& 1.63$\times10^{ -10}$			\\
5.0		& 0.0		&  3.45$\times10^{ -11}$	& 60.9		& 61.0			& 0.0							& 1.0$\times10^{ -30}$			\\
5.0		& 0.5		&  1.39$\times10^{ -11}$	& 60.8		& 61.0			& 0.0							& 3.53$\times10^{ -22}$			\\
5.0		& 1.0		&  9.15$\times10^{ -13}$	& 75.3		& 75.8			& 3.50$\times10^{8}$			& 1.55$\times10^{ -17}$			\\
5.0		& 1.5		&  2.09$\times10^{ -14}$	& 145.9		& 141.5			& 4.06$\times10^{11}$			& 1.88$\times10^{ -14}$			\\
5.0		& 2.0		&  1.26$\times10^{ -16}$	& 820.5		& 164.9			& 2.54$\times10^{13}$			& 5.854$\times10^{ -12}$			\\
10.0		& 0.0		&  7.26$\times10^{ -12}$	& 43.9		& 44.0			& 0.0							& 7.66$\times10^{ -27}$			\\
10.0		& 1.0		&  3.48$\times10^{ -12}$	& 44.4		& 44.3			& 0.0							& 2.91$\times10^{ -22}$			\\
10.0		& 2.0		& 3.84$\times10^{ -13}$	& 52.9		& 52.9			& 2.41$\times10^{7}$			& 2.73$\times10^{ -18}$			\\
10.0		& 3.0		& 1.80$\times10^{ -14}$	& 90.2		& 88.5			& 3.58$\times10^{10}$			& 1.23$\times10^{ -15}$			\\
10.0		& 4.0		& 2.86$\times10^{ -16}$	& 259.1		& 121.1			& 5.76$\times10^{12}$			& 7.40$\times10^{ -13}$			\\
20.0		& 0.0		&  1.45$\times10^{ -12}$	& 26.1		& 26.1			& 0.0							& 2.41$\times10^{ -25}$			\\
20.0		& 2.0		& 8.00$\times10^{ -13}$	& 27.3		& 28.0			& 0.0							& 1.60$\times10^{ -21}$			\\
20.0		& 4.0		& 1.86$\times10^{ -13}$	& 32.3		& 34.0 			& 9.25$\times10^{4}$			& 2.82$\times10^{ -19}$			\\
20.0		& 6.0		&1.12$\times10^{ -14}$	& 53.7		& 56.0			& 4.58$\times10^{9}$			& 7.91$\times10^{ -17}$			\\
20.0		& 8.0		& 3.92$\times10^{ -16}$	& 117.4		& 90.2			& 7.90$\times10^{11}$			& 3.95$\times10^{ -14}$			\\
30.0		&0.0		&5.43$\times10^{-13}$	&19.0		&19.2			&0.0							&4.50$\times10^{-23}$			\\
30.0		&3.0		&3.21$\times10^{-13}$	&28.8		&29.2			&5.37						&1.05$\times10^{-20}$			\\
30.0 		&6.0		&8.84$\times10^{-14}$	&37.4		&38.2			&1.07$\times10^{7}$				&2.73$\times10^{-19}$			\\
30.0		&9.0		&7.38$\times10^{-15}$	&53.6		&55.2			&2.21$\times10^{9}$				&2.21$\times10^{-17}$			\\
30.0		&12.0	&3.80$\times10^{-16}$	&101.4		&87.9			&2.48$\times10^{11}$			&8.27$\times10^{-15}$			\\
40.0		&0.0		&2.62$\times10^{-13}$	&16.3		&16.3			&2.51$\times10^{-29}$			&1.01$\times10^{-22}$			\\
40.0		&4.0		&1.62$\times10^{-13}$	&27.4		&27.3			&1.86$\times10^{2}$				&9.96$\times10^{-21}$			\\
40.0		&8.0		&4.96$\times10^{-14}$	&34.2		&34.8			&9.97$\times10^{6}$				&1.80$\times10^{-19}$			\\
40.0		&12.0	&5.10$\times10^{-15}$	&50.6		&50.3			&1.25$\times10^{9}$				&9.75$\times10^{-18	}$			\\
40.0		&16.0	&3.37$\times10^{-16}$	&87.3		&79.5			&1.09$\times10^{11}$			&3.20$\times10^{-15}$			\\
50.0		&0.0		&1.42$\times10^{-13}$	&14.5		&14.6			&4.41$\times10^{-19}$			&2.51$\times10^{-22}$			\\
50.0		&5.0		&9.07$\times10^{-14	}$	&26.9		&25.9			&2.43$\times10^{3}$				&1.09$\times10^{-20	}$			\\
50.0		&10.0	&3.02$\times10^{-14}$	&32.2		&32.6			&1.48E$\times10^{7}$			&1.19$\times10^{-19}$			\\
50.0		&15.0	&5.18$\times10^{-15	}$	&44.2		&43.4			&5.16$\times10^{8}$				&1.64$\times10^{-18	}$			\\
50.0		&20.0	&4.59$\times10^{-16	}$	&74.0		&70.5			&2.00$\times10^{10}$			&7.23$\times10^{-16	}$			\\
\end{tabular}
\tablecomments{
These are the parameters used for the 35 points used to represent the different radial distances and vertical heights in the model. Note the gas temperature is capped at 4200 K. 
}
\end{center}
\end{table}\label{points}

\end{document}